\documentclass[a4paper,fleqn]{cas-dc}
\pdfoutput =1
\usepackage[authoryear,longnamesfirst]{natbib}
\usepackage{threeparttable}
\usepackage[ruled]{algorithm2e}
\usepackage{pifont}
\usepackage{subfig}
\usepackage{graphicx}
\def\tsc#1{\csdef{#1}{\textsc{\lowercase{#1}}\xspace}}
\tsc{WGM}
\tsc{QE}
\begin{document}
\let\WriteBookmarks\relax
\def\floatpagepagefraction{1}
\def\textpagefraction{.001}
\let\printorcid\relax

\shorttitle{Shorter Latency of Real-time Epileptic Seizure Detection via Probabilistic Prediction}
% \shorttitle{Deep Learning-based Probabilistic Classification for Epileptic Seizure Onset Detection to Shorten Detection latency}
\shortauthors{Yankun Xu, Jie Yang, et al.}  

% \title [mode = title]{Deep Learning for Short-Latency Epileptic Seizure Detection with Probabilistic Classification}
\title [mode = title]{Shorter Latency of Real-time Epileptic Seizure Detection via Probabilistic Prediction}  

% \tnotemark[<tnote number>] 


\author[1]{Yankun Xu}
\author[1]{Jie Yang}
\fnmark[*]
\author[2]{Wenjie Ming}
\author[2]{Shuang Wang}
\author[1]{Mohamad Sawan}
\fnmark[*]

\affiliation[1]{organization={Center of Excellence in Biomedical Research on Advanced
Integrated-on-chips Neurotechnologies (CenBRAIN), School of Engineering, Westlake University},
            city={Hangzhou},
            postcode={310024}, 
            country={China}}

\affiliation[2]{organization={Epilepsy Center, Department of Neurology, Second Affiliated Hospital, School of Medicine},
            city={Hangzhou},
            postcode={310024},
            country={China}}

\cortext[1]{Corresponding authors: Jie Yang, Mohamad Sawan}
\cortext[1]{Email address in order: xuyankun@westlake.edu.cn, yangjie@westlake.edu.cn, hflmwj@zju.edu.cn, wangs77@zju.edu.cn, sawan@westlake.edu.cn}

\begin{abstract}
Although recent studies have proposed seizure detection algorithms with good sensitivity performance, there is a remained challenge that they were hard to achieve significantly short detection latency in real-time scenarios. In this manuscript, we propose a novel deep learning framework intended for shortening epileptic seizure detection latency via probabilistic prediction. We are the first to convert the seizure detection task from traditional binary classification to probabilistic prediction by introducing a crossing period from seizure-oriented EEG recording and proposing a labeling rule using soft-label for crossing period samples. And, a novel multiscale STFT-based feature extraction method combined with 3D-CNN architecture is proposed to accurately capture predictive probabilities of samples. Furthermore, we also propose rectified weighting strategy to enhance predictive probabilities, and accumulative decision-making rule to achieve significantly shorter detection latency. We implement the proposed framework on two prevalent datasets -- CHB-MIT scalp EEG dataset and SWEC-ETHZ intracranial EEG dataset in patient-specific leave-one-seizure-out cross-validation scheme. Eventually, the proposed algorithm successfully detected 94 out of 99 seizures during crossing period and 100\% seizures detected after EEG onset, averaged 14.84\% rectified predictive ictal probability (RPIP) errors of crossing samples, 2.3 s detection latency, 0.08/h false detection rate (FDR) on CHB-MIT dataset. Meanwhile, 84 out of 89 detected seizures during crossing period, 100\% detected seizures after EEG onset, 16.17\% RPIP errors, 4.7 s detection latency, and 0.08/h FDR are achieved on SWEC-ETHZ dataset. The obtained detection latencies are at least 50\% shorter than state-of-the-art results reported in previous studies.

\end{abstract}


% \begin{highlights}
% \item 
% \item 
% \item 
% \end{highlights}

% Keywords
\begin{keywords}
\sep Epilepsy
\sep Seizure detection
\sep EEG
\sep Deep Learning
\sep Probabilistic prediction
\sep Brain-computer interface
\end{keywords}

\maketitle

% Main text
\section{Introduction}
Epileptic patients suffering from unprovoked recurrent seizures occupy approximately 1\% population worldwide (\cite{KISSANI2020106854,xu2022trends}). Seizure is originated from abnormally discharged neurons inside a small brain region, then discharging current is spread to other regions. Two main seizure types, named focal and generalized seizure, depend on seizures that begin in one area then spread to other areas or seizures that begin throughout the brain cortex simultaneously. Long-term drug therapy is one of the major treatments intended for epileptic patients, however, one-third of patients face drug-resistant epilepsy (\cite{RAMGOPAL2014291,assi2017towards}).
Recently, brain-computer interface (BCI) has been broadly applied to healthcare domains (\cite{9328561,ZHUANG202012,VILELA202087,s18103342}). A BCI-based closed-loop seizure detection system that consists of recording, detection, and stimulation elements, can heavily support epileptic patients (\cite{9173735}). Especially for patients with tonic-clonic seizures because electrical stimulation interventions can be operated in time prior to the beginning of convulsions.

\begin{figure}[!t]
\centering
\includegraphics[width=\columnwidth]{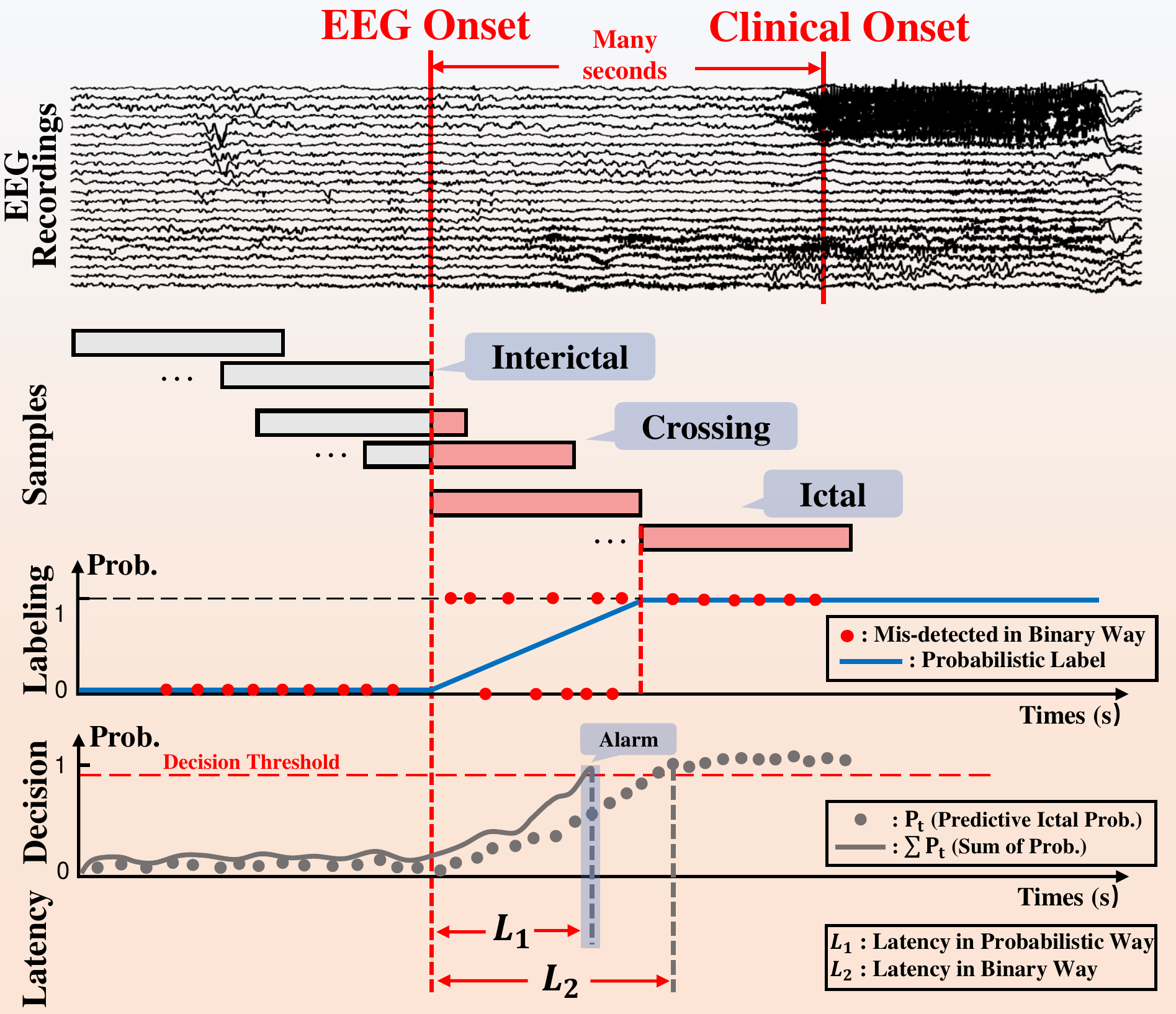}
\caption{Schematic figure for illustrating EEG recordings, segmented samples, traditional seizure detection challenges, and expected decision-making system.}
\label{onset}
\end{figure}

As typical BCI monitoring modalities, scalp electroencephalogram (EEG) and intracranial EEG (iEEG) are widely applied to supervise epileptic patients by recording brain activities. Usually, patients suffer from a couple of seizures per day, and seizure onsets and endings are identified by an experienced epileptologist. Clinically the period between seizure onset and seizure end is defined as ictal period usually lasting 30 s to 2 min, followed by a postictal period usually lasting 5-30 min (\cite{scher1993,Fisher2014,fisher2010definition}). The other period appearing in the health state is defined as interictal period. Medical experts annotate the EEG onset time according to the aberrant signs occurred in the EEG recordings from epileptic patients. However, patients do not behave abnormally at EEG onset time immediately, usually unequivocal EEG onset precedes clinical onset by several seconds. Litt et al. (\cite{LITT200151, LITT200222}) announced this gap would be 7-10 s. The clinical onset refers to the appearance of relevant symptoms, such as convulsion and jerking, that can be obviously reflected in the EEG recordings. 
Latency refers to the delay between EEG onset marked by experts and detected seizure by detection system. Hence, an accurate real-time epileptic seizure onset detection algorithm with shorter latency is becoming a benefit for the patients (\cite{ulate2016automated}). As shown in the top of Fig. \ref{onset}, \textit{EEG Recordings} part displays a real EEG recording example of a patient around the time of EEG onset.

\begin{figure}[!t]
\centerline{\includegraphics[width=\columnwidth]{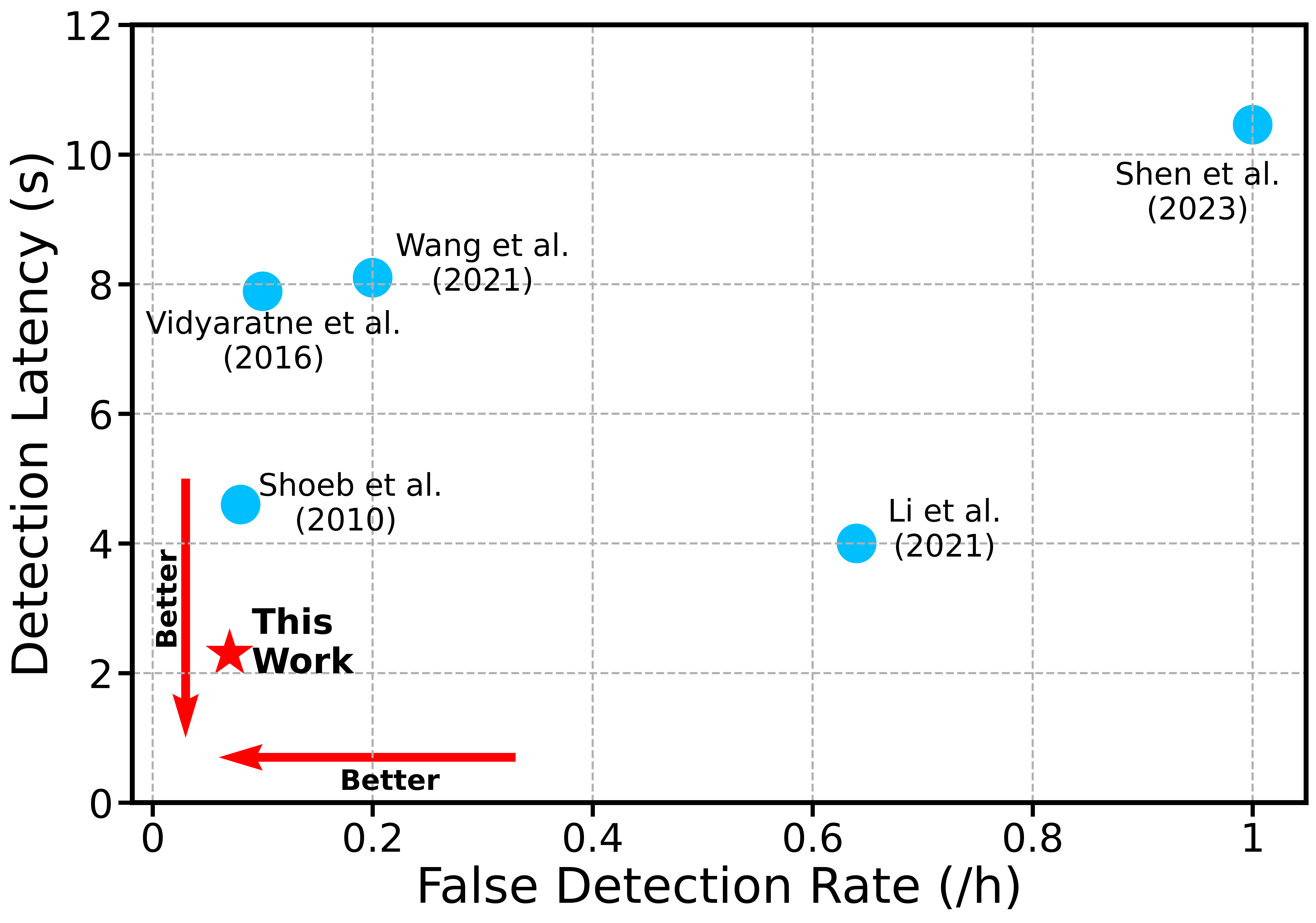}}
\caption{Performance comparisons of this work and selected prior-art studies on CHB-MIT dataset. Shorter detection latency and lower false detection rate simultaneously means better performance.}
\label{comparePer}
\end{figure}

Over the past decades, numerous algorithm-based seizure detection studies have been published, and most of works announced they achieved high sensitivity and low false detection rate (FDR) at the same time (\cite{shoeibi2022overview}). However, high sensitivity alone is still far away from actual seizure intervention usage. Because in terms of real epileptic patient supervision scenarios, short enough detection latency is crucial to guarantee the risk alarms promptly and interventions can be effectively operated prior to serious onset symptoms. Unfortunately, most previous studies overlooked this important metric.
According to our best knowledge, most previous seizure detection studies trained the machine learning-based or deep learning (DL)-based seizure detection algorithm as a binary classification model to distinguish the segmented interictal and ictal samples extracted from corresponding periods, which is shown in \textit{Samples} part of Fig. \ref{onset}. 
However, this strategy remains a drawback that the trained binary classification model cannot correctly detect the crossing samples consisting of partial interictal and ictal components. As shown in \textit{Labeling} part of Fig. \ref{onset}, interictal and ictal samples are labeled as 0 or 1 in traditional binary way for training and trained binary model can recognize them as 0 or 1 correctly, but trained binary model would wrongly detect the crossing samples as 0 or 1 randomly according to our experiments. The reason is that crossing samples are significantly different from the major part of ictal period because they are close to the interictal period, if crossing samples are directly considered as ictal periods in binary way, the binary classification model would only learn the ictal samples with obvious oscillation characteristics mainly instead of crossing samples. It should be noted that the time of each detected dot in the figure is consistent with the end of detected samples.

An accurate and prompt seizure detection system is expected to detect crossing samples in linearly increasing probabilistic format according to the corresponding percentage of ictal component, meanwhile keeping the complete interictal or ictal samples in binary format. Then, an accumulative decision-making rule can be used to alarm the seizure occurrence in short latency. In \textit{Decision} part of Fig. \ref{onset}, gray dots represent predictive probabilities of samples in real-time, and the blue line shows accumulative probability. When the accumulative probability reaches the decision threshold, the detection system would alarm. 
Therefore, as shown in \textit{Latency} part of Fig. \ref{onset}, the expected detection latency $L_{1}$ can be shorter than the length of segmented samples, while the binary classification model can only achieve the latency $L_{2}$ at least longer than the length of segmented samples. 

As mentioned above, detection latency and FDR are two crucial metrics to evaluate the performance of proposed seizure detection algorithms. However, according to previous related studies, many researchers overlooked these two metrics. In Fig. \ref{comparePer}, we compare our work with prior-art studies reporting detection latency and FDR. Shorter detection latency and lower false detection rate simultaneously are used to assess the advantages of algorithms, so that this figure can show the merit of our work.

In this manuscript, we propose a novel DL-based framework intended for shortening seizure onset detection latency via probabilistic prediction. The main innovative contributions are as follows:
\begin{itemize}
\item We are the first to introduce crossing period and convert the seizure detection task from traditional binary classification to probabilistic prediction.
\item We propose a novel DL model that is multiscale STFT-based 3D convolutional neural networks (M-3D-CNN) to accurately capture predictive probabilities of samples.
\item A rectified probability weighting strategy is proposed to further enhance the probabilistic results. 
\item And an accumulative decision-making rule is proposed to achieve shorter latency and lower FDR simultaneously.
\end{itemize}

The remaining content of this manuscript is organized as: We describe in Section 2 recent related works about the seizure detection field. Section 3 elaborates on utilized materials about dataset, data processing, and how to introduce crossing period from seizure-related EEG recordings.
Section 4 carefully illustrates the proposed framework that implements short-lantecy seizure detection. Section 5 clarifies experimental settings and achieved performance. Sections 6 and 7 are the objects of discussion and conclusion, respectively.

\section{Related works}

Over the past decades, algorithm-based seizure detection study is a hot topic attracted by numerous researchers. Shoeb et al. (\cite{SHOEB2004483}) initially detected 131 of 139 seizure events with 8 s detection latency and 0.25/h FDR on 36 clinical pediatric subjects in 2004. Then they published significant performance that 96\% detection sensitivity with averaged 4.6 s latency on 24 pediatric subjects in 2010 (\cite{ShoebG10}). They took advantage of spectral and spatial features combined with support vector machine (SVM) classifier to achieve advanced results. As pioneers of this field, Shoeb et al. (\cite{ShoebG10}) also published CHB-MIT scalp dataset, it has become one of the most famous and important resource intended for seizure-related study. In 2011, from the same group, researchers applied similar approaches to the clinical iEEG dataset from 10 patients and achieved 97\% detection sensitivity, 0.03/h FDR, and 5 s detection latency (\cite{KHARBOUCH2011S29}).

In 2016, Vidyaratne et al. (\cite{7727334}) used recurrent neural network to process raw EEG signals, and achieved 100\% sensitivity, 0.08/h FDR, and 7 s detection latency on CHB-MIT dataset. 
In (\cite{7927402}), authors applied statistical and morphological features combined with an adaptive distance-based change point detector to achieve 96\% sensitivity, 0.12/h FDR, and 4.21 detection latency, respectively on CHB-MIT dataset.
The empirical mode decomposition method is a prevalent technique broadly applied to seizure detection applications, numerous authors utilized it and its variation to achieve satisfactory performance over the past decade (\cite{HASSAN2020105333,8368709,9339990}).
The short-time Fourier transform (STFT) is an effective method widely used to extract seizure-related features. Yuan et al. \cite{8470079} proposed a multi-view DL framework for EEG seizure detection based on STFT and convolutional neural networks (CNN), they achieved 94.37\% accuracy using 5-fold patient-specific cross-validation on CHB-MIT dataset. 
And in (\cite{SHOEIBI2021113788}), authors compared handcrafted features and convolutional autoencoders for seizure detection performance, and showed Butterworth filter, Different Features and Fisher method are better.
Different from the traditional convolution operation that considers EEG signal or STFT features as 2D image-like features, 1D-CNN architecture is also applied to seizure-related studies to (\cite{10.1145/3241056,WANG2021212,9767764,JANA2020403}). In (\cite{WANG2021212}), authors achieved sensitivity, FDR, and detection latency of 99.31\%, 0.2/h, and 8.1 s from event-based level on CHB-MIT dataset. Li et al. (\cite{8995501}) proposed a novel channel-embedding spectral-temporal squeeze-and-excitation network using wavelet features to recognize epileptic EEG signals, they achieved 92.41\% sensitivity and 96.05\% specificity on CHB-MIT dataset. In (\cite{AGHABABAEI2021114630}), authors provided a orthogonal matching pursuit algorithm aiming for data compression, and achieved 0.94 AUC performance. Zhou et al. (\cite{ZHOU2023119613}) combined phase space reconstruction, Fisher discriminant analysis and minimum distance to the Riemannian means algorithm for achieving 96.01\% sensitivity, 98.32\% specificity on CHB-MIT dataset.

Burrello et al. (\cite{8584751,8715186, 8723166}) from Swiss research group published a seizure-oriented iEEG database, known as SWEC-ETHZ database. They achieved 94.84\% specificity and 95.42\% accuracy on short-term dataset, and detected 79 out of 92 unseen seizures without any false alarms across all the patients on long-term dataset. 
Afterwards, Wang et al. (\cite{WANG2021212}) and Sun et al. (\cite{9858598}) obtained sensitivity, FDR, detection latency of 97.52\%, 0.07/h FDR, 13.2 s and 97.5\%, 0.06/h FDR, 13.7 s, respectively.

From our point of view, there is a remained controversy in measuring detection latency metric. As we depicted in Fig. 1, binary classification model cannot achieve the latency shorter than the length of segmented samples. According to prior-art publications, they segmented EEG samples in at least 5 s, however, most of them announced their proposed algorithms can achieve less than 5 s detection latency. Hence, we doubt that previous researchers overlooked the crossing samples from real-time perspectives, leading to compute the detection latency by measuring the distance between EEG onset and begin of detected sample instead of end of detected sample. In this manuscript, we pioneer introducing crossing period and probabilistic prediction in real-time seizure detection task to achieve short detection latency.

\section{Materials}
\subsection{Dataset}
In this work, EEG and iEEG datasets are both considered to validate the proposed framework. The CHB-MIT scalp EEG dataset is one of the most prevalent open access datasets intended for seizure-related research. There are 23 pediatric patients monitored by 22 electrode channels with 256 Hz sampling rate, and each subject obtains various numbers of seizure and non-seizure EEG recording files in 1h. The earliest change associated with the clinical seizure is annotated as EEG onset by clinical experts. We ignored the subjects with changing electrode placement, thereby 19 subjects are selected for the following experiments. 

The SWEC-ETHZ dataset is an emerging and open-access iEEG dataset collected from pre-surgical evaluations of patients with pharmacoresistant epilepsies, it was published in 2018. This database contains two versions according to different recording durations -- long-term and short-term versions. In this work, we use short-term version to test our proposed algorithm. In terms of short-term dataset, each patient obtains several seizure files, and each file consists of a 3-min interictal period immediately followed by an ictal period and a 3-min postictal period. The EEG onset time is identified by an experienced epileptologist. Because there is insufficient interictal duration for model training on short-term dataset, we only select patients with no less than 4 seizures in this dataset, so that 11 of 16 patients are selected. Furthermore, the number of implanted electrode channels used in these patients varies from 42 to 100, and 512 Hz sampling rate is chosen. Table 1 summarizes the characteristics of two datasets used in this work.

\subsection{Data preparation}
Furthermore, duration of interictal period is much longer than duration of ictal period on CHB-MIT dataset, so that we overcome this unbalanced data issue by extracting interictal samples without any overlaps, and ictal samples with 80\% overlaps, respectively. And we data-point-wisely extracted crossing samples. Because the duration of crossing period for each seizure equals to the length of segmented samples, the duration of crossing period should be 5 s, and the number of extracted crossing samples for each seizure is computed by duration of segmented samples multiplying sampling rate (5 (s) $\times$ 256 (Hz) = 1280 (samples)). As for SWEC-ETHZ dataset, we directly extracted samples of all three periods with 80\% overlaps.

\subsection{Definitions of seizure-related periods}
DL algorithms cannot train the model directly on the successive EEG recordings, so that we need to extract successive segmented samples from the recording in advance. Firstly, we need to identify the interictal, ictal and crossing periods. According to the seizure file of CHB-MIT dataset, beginning and ending time of each seizure is provided, and we manually set a 30 min postictal period following seizure ending, then left part is belong to interictal period. In terms of short-term version of SWEC-ETHZ dataset, each patient obtains several seizure recording files, each file consists of a 3 min interictal segment immediately followed by an ictal segment and a 3min postictal segment. 

\begin{figure}[!t]
\centering
\includegraphics[width=\columnwidth]{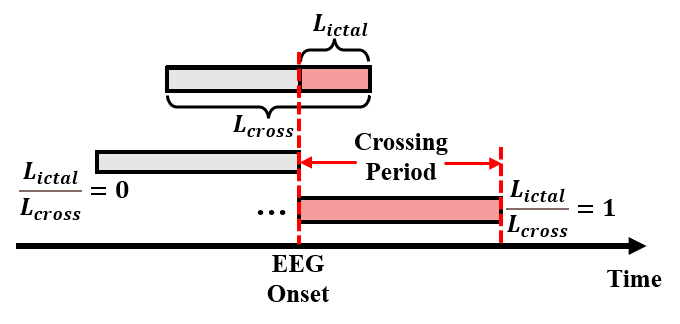}
\caption{Explanation of crossing period. In crossing period, extracted sample consists of partial interictal and ictal component simultaneously. The duration of crossing period depends on the length of segmented samples.}
\label{crossing}
\end{figure}

The schematic figure for definition of crossing period is shown as Fig. \ref{crossing}, where $\frac{L_{ictal}}{L_{cross}}$ denotes the length of ictal component occupying the whole length of crossing sample. The crossing period is defined as ending of extracted samples begins at EEG onset time (the last time for $\frac{L_{ictal}}{L_{cross}}=0$) and ends at a duration of extracted sample after EEG onset (the first time for $\frac{L_{ictal}}{L_{cross}}=1$). In experiments, samples are extracted in 5 s and 10 s segments for CHB-MIT dataset and SWEC-ETHZ dataset, respectively.

\begin{table}[!t]
    \scriptsize
    \caption{Summary of two datasets used in this work.}
    \centering
    \renewcommand\arraystretch{1}
        \begin{tabular}[]{c|c|c|c|c|c}
        \toprule
        \makecell{Dataset} & \makecell{EEG \\ type} & \makecell{\# of \\ selected \\ patients} & \makecell{\# of \\ channels} & \makecell{\# of \\ seizures} & \makecell{Interictal \\ duration}\\
        \midrule
        \makecell{CHB-MIT} & \makecell{EEG} & \makecell{19} & \makecell{22} & \makecell{99} & \makecell{335.5h} \\
        \makecell{SWEC-ETHZ } & \makecell{iEEG} & \makecell{11} & \makecell{42$\sim$100} & \makecell{89} & \makecell{4.45h} \\

        \bottomrule
        \end{tabular}
        \begin{tablenotes}
            \scriptsize
            \item[1] iEEG = intracranial EEG
            
        \end{tablenotes}
        % \quad
        
        % \leftline{\text{\scriptsize{\quad sEEG: scalp EEG \quad iEEG: intracranial EEG}}}
        
\end{table}

\section{Proposed framework}
\subsection{Multiscale STFT-based 3D-CNN architecture}

\begin{figure*}[!ht]
\centerline{\includegraphics[width=\textwidth]{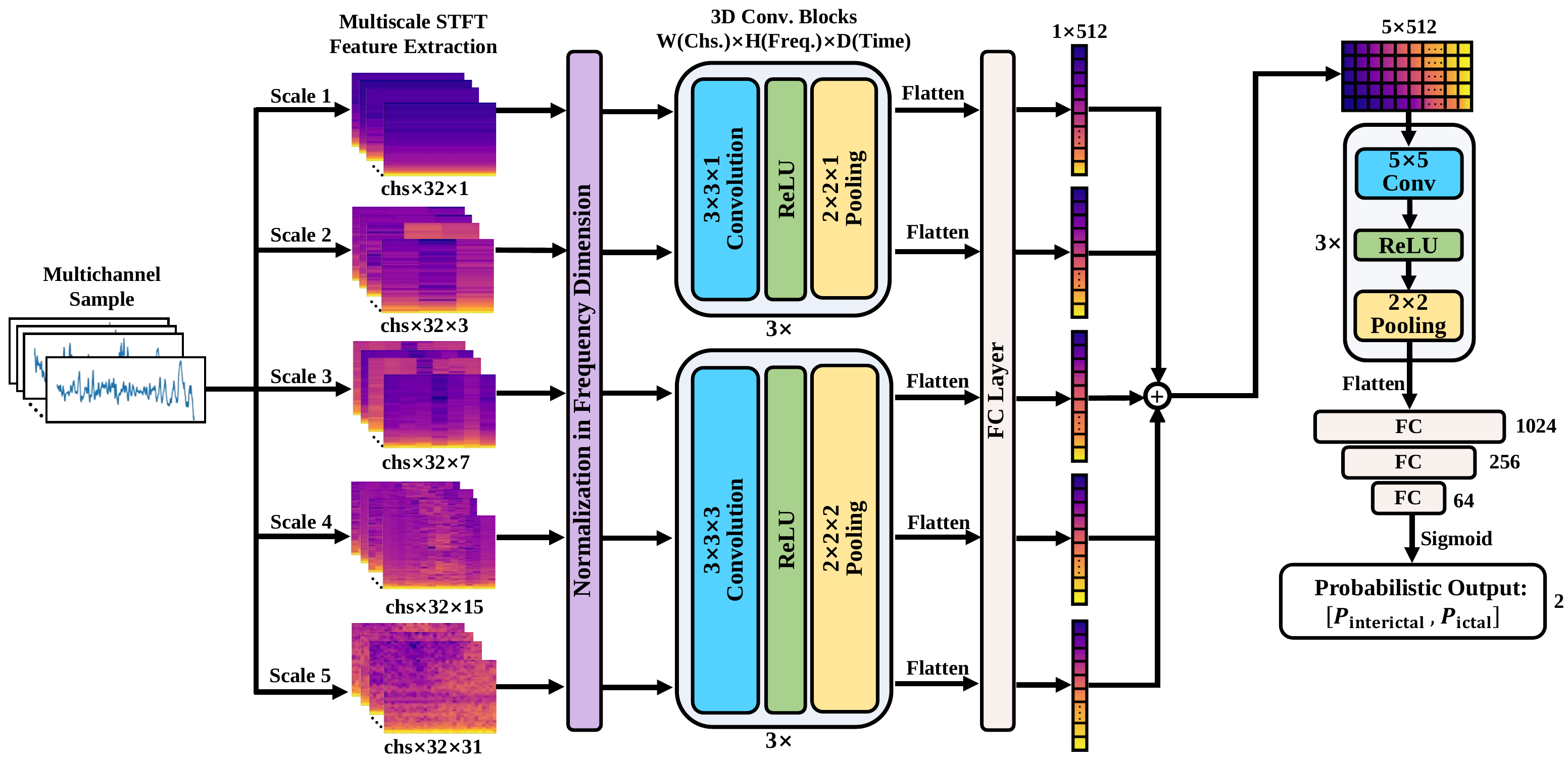}}
\caption{Architecture of proposed multiscale STFT-based 3D convolutional neural networks. The input is a multichannel EEG sample, we extract short-time Fourier transform (STFT) features of the input at scale 1 to 5. Then a FreqNorm layer is used to channel-wisely normalize 3D STFT values from 0 to 1 in frequency dimension at each time step. The 3D STFT feature is fed to 3$\times$ same 3D convolution (Conv.) blocks comprising of 3D Conv., ReLU activation function, and 3D max-pooling. The last layer generated by Conv. block is flattened and connected to a fully connected (FC) layer with 512 nodes. 5 vectors obtained from 5 different scales are concatenated to a 5$\times$512 matrix, then 3$\times$ traditional 2D Conv. blocks and 3$\times$ FC layers with ReLU are used to make classification. The last output FC layer with 2 nodes utilizes a sigmoid activation function to guarantee the output in probability ranged from 0 to 1.}
\label{net}
\end{figure*}

\begin{figure}[!t]
\centerline{\includegraphics[width=\columnwidth]{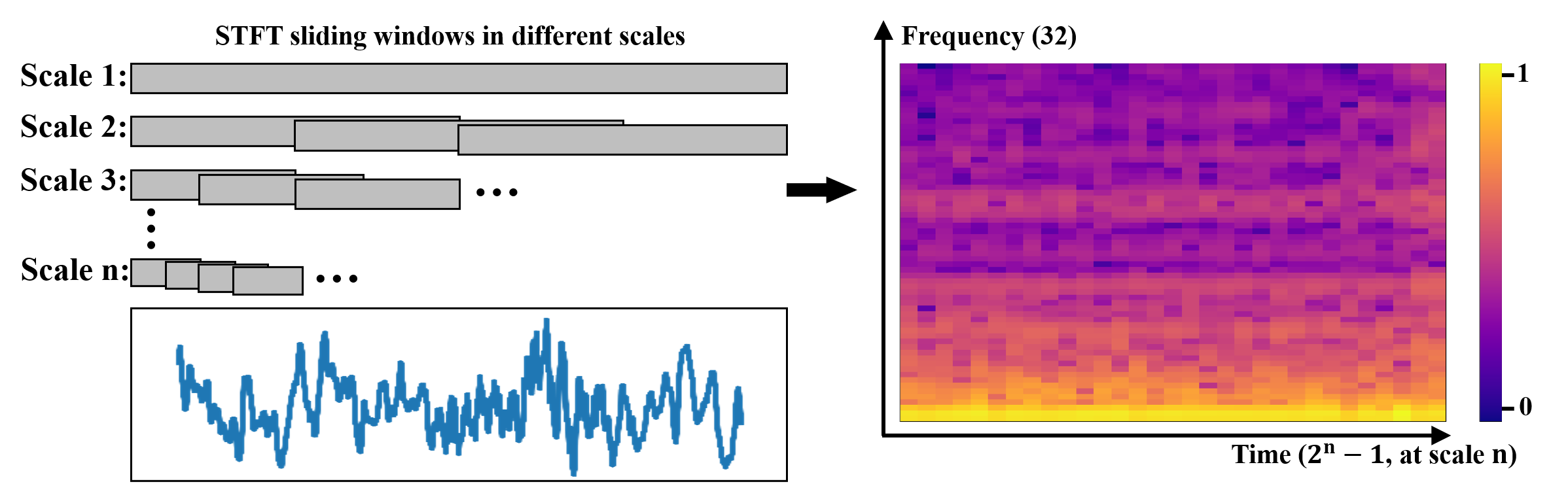}}
\caption{Multiscale STFT feature extraction scheme for single-channel sample}
\label{stft}
\end{figure}

\begin{figure}[!t]
\centerline{\includegraphics[width=\columnwidth]{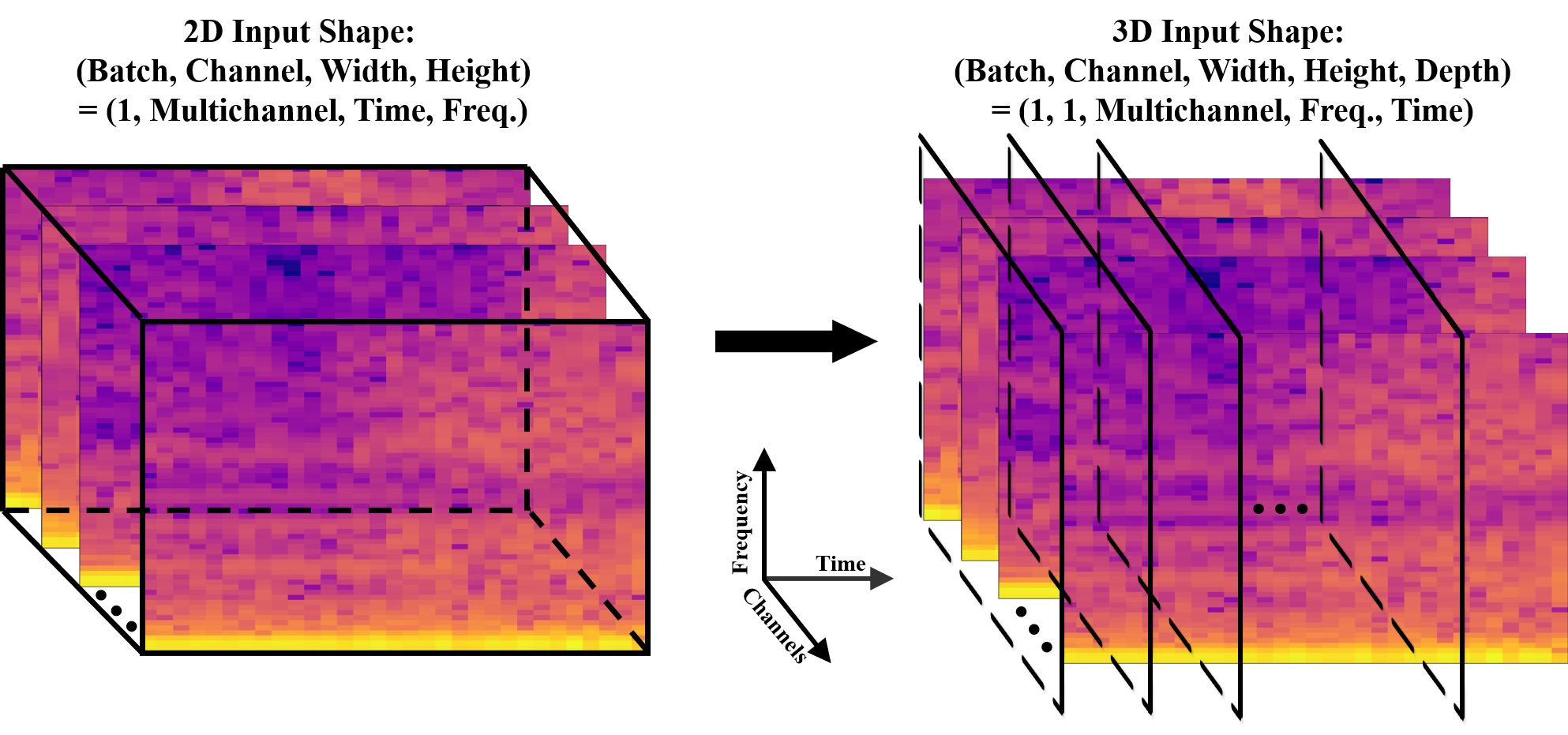}}
\caption{Schematic figure for difference between traditional 2D and proposed 3D convolution operation for multichannel STFT feature at each scale.}
\label{3d}
\end{figure}

As a multiscale architecture, proposed model aims to address the challenge that recognition of target samples in probabilistic way. Fig. \ref{net} shows the detailed architecture of M-3D-CNN model. Each segmented sample input is a period of multichannel EEG signals, we implement multiscale STFT feature extraction at first. As a prevalent and effective signal analysis tool in frequency domain, STFT method is widely used in seizure detection studies (\cite{6909003,8467308}). Although multiscale STFT has been proposed to solve challenges existed in the field of audio and image processing (\cite{juillerat2008enhancing,vyas2018multiscale}), previous studies did not show advantages of this technique and there is no work applying it to the physiological signals yet. Different from previous works, we propose to extract channel-wise STFT features in different scales simultaneously. This feature extraction approach which is in combination with proposed 3D-CNN architecture has been shown to outperform other DL models in this work. The computation way of multiscale STFT is elaborated as follows:

\begin{equation}
\begin{aligned}
X_{scale}[\omega, m] = & \sum\limits^{WL-1}\limits_{k=0} w[k] \cdot x[m\times\frac{WL}{2} + k] \cdot e^{-j2\pi\omega k} \\
WL = & \ \frac{L_{signal}}{scale}, scale = 1,2,3,...\\
\end{aligned}
\end{equation}
where $m$ and $\omega$ are the index of the sliding window and frequency coefficient, respectively. $WL$ and $L_{signal}$ denote window length and length of EEG signal, and $\frac{WL}{2}$ refers to the overlapping length is set to half of window length.

Various scales of STFT in M-3D-CNN model stand for the chosen length of sliding window in STFT indeed, meanwhile we set 50\% overlapping for sliding window, thus the number of detecting windows in time axis of STFT result is $N_{window} = 2^{n}-1$ at scale $n$. Meanwhile, the size of fast Fourier transform (FFT) is set to 64, so that 32 informative coefficients are left in the frequency dimension. Fig. \ref{stft} displays STFT result of the single-channel signal at different scales.
In M-3D-CNN, we select 5 different scales from 1 to 5, because we aim to extract features from small to large scale, and 5th scale is large enough for most real-time situations.
Each scale would generate a 3D STFT feature tensor (channel $\times$ frequency $\times$ time). Previously, most studies considered this type of feature as a 2D image in the shape of H (time) $\times$ W (frequency) with depth whose size is equal to the number of channels, then implemented 2D convolution operation (\cite{ijerph18115780,ahmad2022eeg}). It is an intuitive operation, however, this way cannot take time step information into account. In M-3D-CNN model, we consider 3D STFT feature as a 2D image in the shape of H (channel) $\times$ W (frequency) with time step Depth, then 3D convolution operation is utilized. Fig. \ref{3d} shows the difference between 2D and 3D convolution operations for STFT features. The values from achieved 3D STFT result need to be channel-wisely normalized from 0 to 1 at each time step.

Then, extracted 3D features in different scales are fed to 3$\times$ same 3D convolution blocks, and each block contains a 3D convolution module with kernel size in the shape of 3 $\times$ 3 $\times$ 1 and a max-pooling module with kernel size in the shape of 2 $\times$ 2 $\times$ 1 for the first two scales, and 3 $\times$ 3 $\times$ 3 convolution kernel and 2 $\times$ 2 $\times$ 2 max-pooling kernel are used for the rest three scales. % 2x2x1 for first two blocks
It is noted that we set different kernel sizes between the first two scales and the rest scales, the reason is that we aim to keep the depth (3 layers) of 3D convolution blocks, however, the time dimensions of first two scales are too short to be suitable for 3 $\times$ 3 $\times$ 3 convolution kernel and 2 $\times$ 2 $\times$ 2 max-pooling kernel.
The output generated after 3$\times$ convolution blocks at each scale is flattened, then a fully connected (FC) layer with 512 nodes is followed to generate a 1D vector with the same shape.
This operation aims to unify the shape, thereby eliminating the effects of inconsistent vector dimension due to input EEG signals containing different numbers of channels. 
These 5 vectors originated from 3D feature tensors at 5 different scales are concatenated together to build a 2D matrix in the shape of 5 $\times$ 512. Then successive 3$\times$ same 2D convolution blocks with 5 $\times$ 5 convolution kernel size and 2 $\times$ 2 max-pooling kernel size, are connected to the 2D matrix. The output is also flattened to a 1D vector, and 3 FC layers with 1024, 256 and 64 nodes are followed to make classification. Eventually, we achieve the output in the last FC layer with 2 nodes for representing expected output format in probability pairs.
It is important to note that except for the last layer Sigmoid activation function is used to generate probabilistic output, ReLU activation function is used in the M-3D-CNN model everywhere else.

% \vspace{-1cm}

This novel architecture is inspired by the fact that probabilistic crossing samples are comprised of short complete interictal and ictal periods. $P_{ictal}$ denotes the percentage of a complete ictal period occupying the crossing sample, rather than uniformly distributed in the crossing sample. However, traditional feature extraction approaches either implement FFT for the whole duration or STFT with a single specific scale, which cannot meet the situation that the probability pairs of crossing samples are dynamically changing in real-time scenario. 
% Furthermore, 5 scales setting is applicable for most situations, because we already set 50\% overlapping for the STFT time window and the duration of segmented samples is usually less than 10 s.

\subsection{Probabilistic labeling rule}
In terms of traditional binary classification, interictal and ictal samples are annotated by $[1, 0]$ and $[0, 1]$ according to the one-hot encoding rule. However, the label of crossing sample should contain probabilistic information rather than simple one-hot information. The intuitive operation is to directly label crossing samples as a single probability as a regression task required, but this type of labeling would cause crossing samples cannot be trained with interictal and ictal samples together. We aim to train the M-3D-CNN model using interictal, ictal, and crossing samples together, because crossing samples are comprised of partial interictal and ictal parts, and M-3D-CNN model is expected to learn the unique crossing samples characteristics from complete interictal and ictal samples. 

In this work, we keep labeling ictal and interictal samples as binary format, and take soft-label strategy to annotate the crossing samples in probabilistic format (\cite{10.1136/amiajnl-2013-001964}).
Compared to traditional regression task training, soft-label annotation replace the output vector with the shape of $(1,)$ (e.g., $P_{ictal} = 0.3,0.6,0.8, ...$ ) with the output vector with the shape of $(1,2)$ (e.g., $[P_{interictal}, P_{ictal}] = [0.7, 0.3], [0.4, 0.6], [0.2, 0.8], ...$). There are several advantages of soft-label strategy compared to the regression training. Firstly, the soft-label strategy makes networks can train the crossing samples, complete interictal and ictal samples simultaneously. Secondly, soft-label enables usage of cross entropy loss function, which takes both interictal and ictal information into account rather than only ictal probability provided with simple mean square error loss function in regression task.

In terms of experimental training setting, we keep labels of interictal and ictal samples in the traditional way ($Label_{interictal} = [1, 0], Label_{ictal} = [0, 1]$) and label crossing samples into 20 probability pairs as following rules:
\begin{equation}
\begin{aligned}
Label_{cross} = & \ [P_{interictal}, P_{ictal}] \\
where, \ P_{ictal} = & \ 0.05p \ \ if \ \ \frac{L_{ictal}}{L_{cross}} \leq 0.05p \\
P_{interictal} = & \ 1 - P_{ictal}, \\
p = & \ 0,1,2,... 19 \\
\end{aligned}
\end{equation}
where, $\frac{L_{ictal}}{L_{cross}}$ ranges from 0 to 1. In terms of loss function utilized for model training, we take binary cross entropy loss function which is defined as:
\begin{equation}
\begin{aligned}
\mathcal{L}(P_{ictal}^{(t)}, \hat{P}_{ictal}^{(t)}) = & -(P_{ictal}^{(t)} \cdot log(\hat{P}_{ictal}^{(t)}) \\
& + (1-P_{ictal}^{(t)}) \cdot log(1-\hat{P}_{ictal}^{(t)}))
\end{aligned}
\end{equation}
where, $\hat{P}_{ictal}$ denotes predictive $P_{ictal}$. A \textit{Sigmoid} activation function is used to scale the $\hat{P}_{ictal}$ from 0 to 1 before computing the loss.

\subsection{Rectified weighting strategy}
According to the \textit{Labeling} part of Fig. 1 and Eq. (1), $P_{ictal}$ of crossing samples is expected to be linearly increased from 0 to 1 along with the linearly increasing percentage of ictal period occupying the whole crossing sample ($\frac{L_{ictal}}{L_{cross}}$). In practical experiments, however, predictive ictal probability (PIP) cannot guarantee to be always accurate even if we can obtain quite small loss. Instant inaccurate PIP can damage performance by increasing FDR or decreasing sensitivity. Thus, we propose a rectified weighting strategy to enhance the predictive ictal probability. The theory of this strategy is that the current PIP at time $t$ is not achieved by the current sample (or segment) only, we consider introducing previously achieved PIPs to mitigate the impact of current PIP. Due to real-time scenario, we store the PIPs every 0.1 s generated from previous 5 s, then we utilize PIPs from previous 5 s, 3 s, and 1 s to fit three linear regression (LR) functions, in order to generate three new PIPs (PIP$_{LR5s}$, PIP$_{LR3s}$, PIP$_{LR1s}$) only based on previous PIPs from different durations instead of current PIP$_{t}$. Eventually, we can achieve rectified PIP at time $t$ (RPIP$_{t}$) that is computed as Eq. (4). The weights $\lambda_{1}, \lambda_{2}, \lambda_{3}, \lambda_{4}$ are experimentally set to adjust the weighting of different PIPs.

In short, rectified weighting strategy aims to enhance the current PIP by rendering it more relevant to previous PIPs, this can help reduce the impact of abnormal PIPs which are might be generated by noises, artifacts, or model limitations.

\begin{equation}
\begin{aligned}
RPIP_{t} = 
{
\left[ \begin{array}{cccc}
\lambda_{1} & \lambda_{2} & \lambda_{3} & \lambda_{4}\\
\end{array} 
\right ]}
\cdot
{
\left[ \begin{array}{c}
PIP_{LR5s} \\
PIP_{LR3s} \\
PIP_{LR1s} \\
PIP_{t}
\end{array} 
\right ]}
\end{aligned}
\end{equation}

\subsection{Decision-making rule}
We do not make decision only based on a single PIP because it is difficult to significantly shorten the detection latency. The reason is that if the decision threshold is high, the detection latency is inevitably longer than the duration of segmented samples, meanwhile FDR would be high if we set a low decision threshold. Therefore, in this work, we also propose an accumulative decision-making rule, whose schematic figure is shown in the \textit{Decision} part of Fig. \ref{onset}. 

\begin{equation}
\begin{aligned}
AP_{t} = \frac{\sum^{t}_{i=t-5} RPIP_{i+1}}{r}  \ (\text{if} \ RPIP_{i+1} > RPIP_{i})
\end{aligned}
\end{equation}

Same as rectified weighting strategy, we store the RPIPs from previous 5 s with detection rate $r$, which means we store the RPIPs in a time step of $\frac{1}{r}$ s. Then we compute the accumulative probability (AP) at current time $t$ ($AP_{t}$) as Eq. (5). In short, we only accumulate increased RPIPs during the period of previous 5 s. And the detection system would alarm at time $t_{d}$ when $AP_{t_d} \geq Thr.$, where $Thr.$ represents the decision threshold. This threshold is determined manually to adjust the performance on sensitivity and false alarm rate (FDR) simultaneously according to different scenario needs. Eventually, \textbf{Algorithm 1} illustrates the detailed decision-making rule of proposed framework intended to detect seizures.

% \leftline{\text{\scriptsize{\quad RPIP: rectified Predictive Ictal Probability \quad AP: Accumulative Probability}}}
\begin{equation}
\begin{aligned}
% errors
\text{Event-based Sensitivity} = \ &N_{DC} / N_{Total} \\
\text{RPIP Errors} = \ & \frac{\sum^{T_{c}-1}_{t=0} \sqrt{(P_{ictal}^{(t)}-\hat{P}_{ictal}^{(t)})^{2}}}{T_{c}}  \\
\text{Detection Latency} = \ & t_{d} - t_{onset} \quad \text{if} \ AP_{t_d} \geq Thr. \\
\text{FDR} = \ & N_{FD} /h
\end{aligned}
\end{equation}

\subsection{Performance metrics}
In this work, there are four metrics - sensitivity, errors, detection latency, and FDR, used to investigate the performance of proposed DL-based framework. Eq. (6) reveals how we compute these four metrics. The sensitivity used in this work is event-based sensitivity which computes the number of seizures detected during the crossing period over the total number of seizures in each patient. RPIP errors are computed as the second equation in Eq. (6), we only consider RPIP errors of crossing samples to figure out the capacity of M-3D-CNN model combined with rectified weighting strategy to recognize crossing samples in probabilistic way, $t$ denotes the sample detected at time $t$ and $T_{c}$ refers to the duration of crossing period. In terms of detection latency, we implement \textbf{Algorithm 1} and mark the time ($t_{d}$) when $AP_{t_{d}}$ larger than and equal to decision threshold ($Thr.$), then compute the detection latency by calculating distance between $t_{d}$ and EEG onset time ($t_{onset}$). The last metric is FDR, we directly count the number of false detection according to \textbf{Algorithm 1} during the interictal period per hour as FDR.

\begin{algorithm}[!t]
\caption{Decision-making rule of proposed framework intended to detect seizures.}\label{alg1}
\textbf{Input:} Sample at time $t$: $S_{t}$. \\
\textbf{Output:} Detection time: $t_{d}$. \\
\textbf{Initialize:} Detection rate: $r$; Decision threshold: $Thr.$. \\
 Stacking vector with zeros for previous 5s RPIPs: $\mathbb{P} = [P_{t-5}, P_{t-5+\frac{1}{r}}, ..., P_{t-\frac{1}{r}}, P_{t}] = \mathbb{0}$. \\
\While{True}{
$[1-PIP_{t}, PIP_{t}] \gets \textit{M-3D-CNN}(S_{t})$\;
$RPIP_{t} \gets RWS(PIP_{t})$\;
$\mathbb{P} \gets append(\mathbb{P}[P_{t-5+\frac{1}{r}}:end],RPIP_{t})$ \\
$AP_{t} \gets \sum^{t}_{i=t-5} \mathbb{P} \ (\text{if} \ RPIP_{i+1} > RPIP_{i})$ \\
\If{$AP_{t_{d}} > Thr.$}{
      Alarm at time $t_{d}$; \\
      Refresh $\mathbb{P} = \mathbb{0}$
    }
}

\scriptsize{Abbr.:}\\
\scriptsize{RPIP: Rectified Predictive Ictal Probability }\\
\scriptsize{M-3D-CNN: Multiscale STFT-based 3D-CNN}\\
\scriptsize{RWS: Rectified Weighting Strategy \quad AP: Accumulative Probability}
\end{algorithm}

\begin{figure*}[!ht]
% 	\centering
	\subfloat[]{\includegraphics[width=0.32\textwidth]{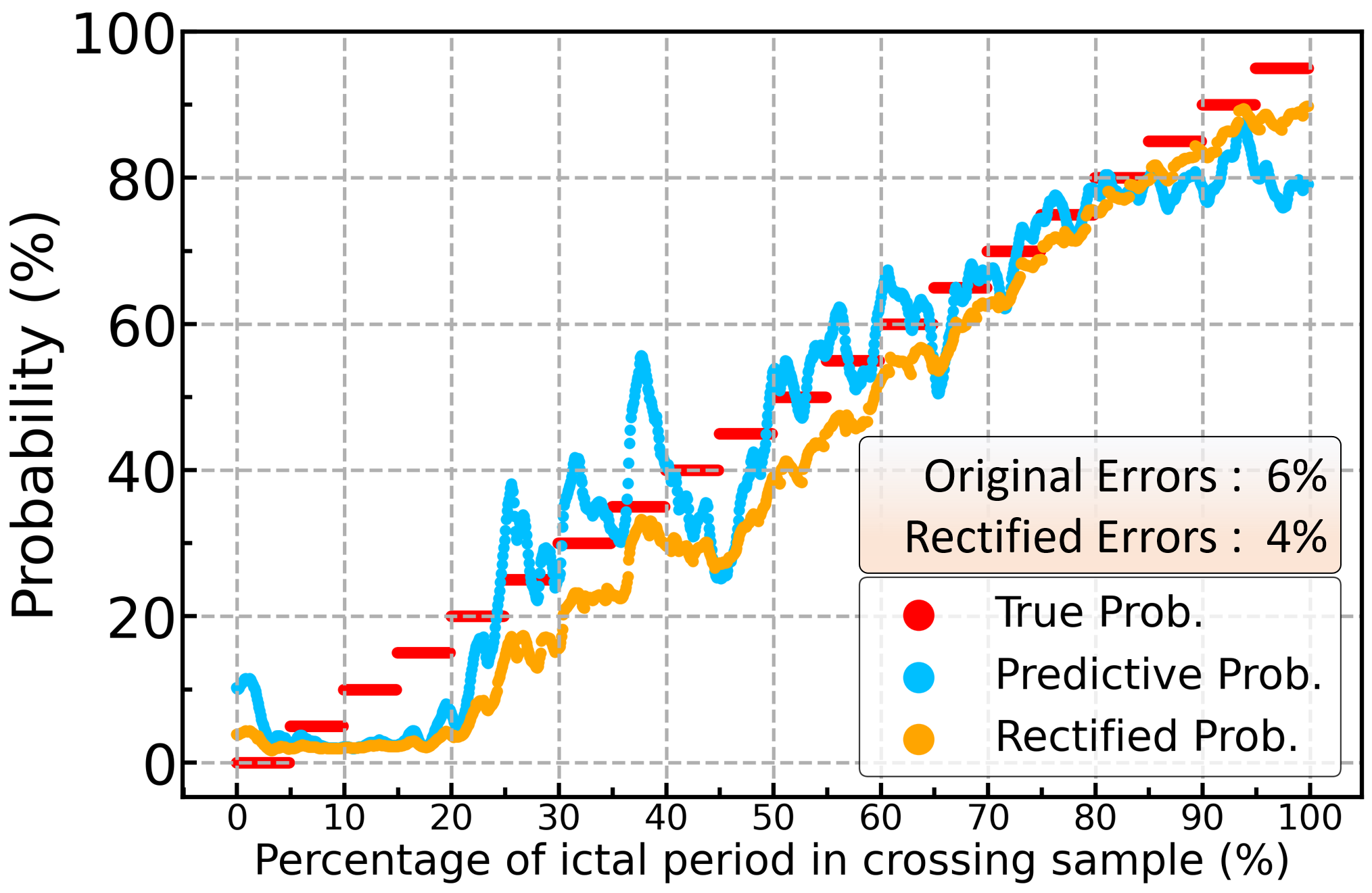}} \hfill 
	\subfloat[]{\includegraphics[width=0.32\textwidth]{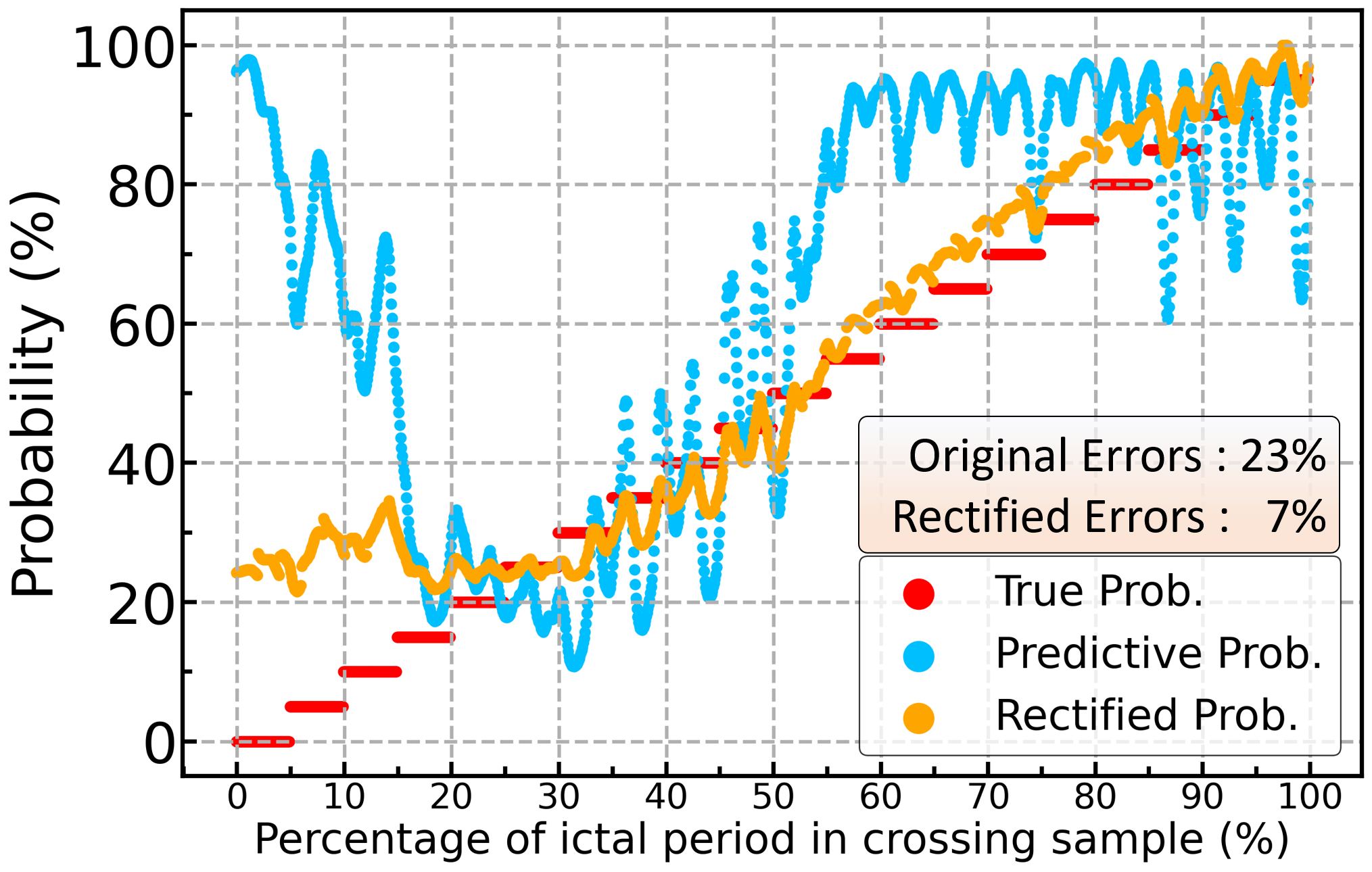}} \hfill 
	\subfloat[]{\includegraphics[width=0.32\textwidth]{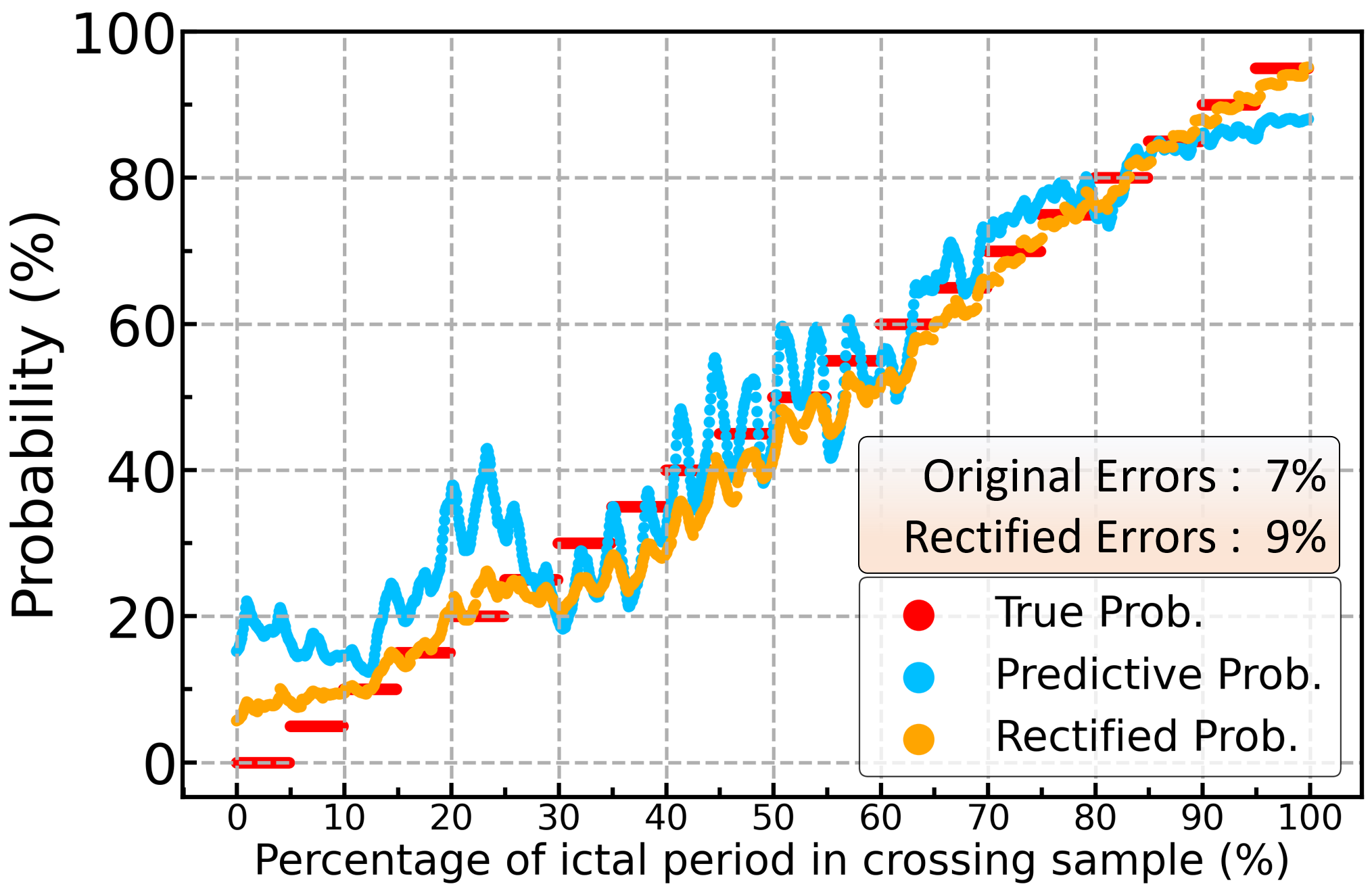}} \hfill 
% 	\subfloat[]{\includegraphics[width=0.25\textwidth]{prob-4.png}} \hfill 
	
% 	\text{\scriptsize{Conv: Convolution, MP: Max pooling, Act: Activate function, FC: Fully connected}}

 	\caption{Performance of rectified probability weighting strategy. Here are 3 representative examples, each figure refers to crossing period of a seizure. Red, blue and orange dots stand for true, predictive, and rectified probabilities, respectively. The original errors achieved by PIPs and rectified errors achieved by RPIPs are also highlighted. }
    
\label{rws}
\end{figure*}

\begin{table*}[!h]
    \scriptsize
    \caption{Performance of proposed algorithm on two datasets in the patient-specific and leave-one-seizure-out cross-validation scheme. Rectified predictive ictal probability (RPIP) is generated by M-3D-CNN model followed by rectified weighting strategy. Detection latency and false detection rate (FDR) are obtained by accumulative decision-making rule. There are 19 and 11 cases in CHB-MIT and SWEC-ETHZ datasets, respectively. We provide mean and standard deviation values from every seizure of each patient for RPIP errors, detection latency, and FDR metrics.}
    \centering
    \renewcommand\arraystretch{1.5}
        \begin{tabular}[]{c|c|c|c|c||c|c|c|c|c}
        \toprule
        \multicolumn{5}{c||}{CHB-MIT dataset} & \multicolumn{5}{c}{SWEC-ETHZ dataset}\\
        \midrule
        \makecell{Patient \\ ID} & \makecell{Sensitivity \\ (N$_{\text{DC}}$ / N$_{\text{T}}$)} & \makecell{RPIP Errors of \\ Crossing Samples \\ (\%)} & \makecell{Detection \\ Latency \\ (s)} & \makecell{FDR \\ (/h)} & \makecell{Patient \\ ID} & \makecell{Sensitivity \\ (N$_{\text{DC}}$ / N$_{\text{T}}$)} & \makecell{RPIP Errors of \\ Crossing Samples \\ (\%)} & \makecell{Detection \\ Latency \\ (s)} & \makecell{FDR \\ (/h)} \\
        \midrule
        \makecell{chb01} & \makecell{7/7} & \makecell{10.85 $\pm$ 6.05} & \makecell{1.9 $\pm$ 0.6} & \makecell{0} & \makecell[l]{\hspace{0.05cm} ID1} & \makecell{13/13} & \makecell{11.18 $\pm$ 8.27} & \makecell{3.6 $\pm$ 1.0} & \makecell{0.02 $\pm$ 0.04}\\
        
        \makecell{chb02} & \makecell{2/3} & \makecell{\hspace{0.04cm} 29.76 $\pm$ 12.49} & \makecell{2.9 $\pm$ 1.8} & \makecell{0.42 $\pm$ 0.36} & \makecell[l]{\hspace{0.05cm} ID2} & \makecell{4/4} & \makecell{17.90 $\pm$ 6.98} & \makecell{3.8 $\pm$ 1.5} & \makecell{0.19 $\pm$ 0.26} \\
        %\hline
        \makecell{chb03} & \makecell{7/7} & \makecell{\hspace{0.04cm} 9.10 $\pm$ 2.74} & \makecell{2.3 $\pm$ 0.4} & \makecell{0} & \makecell[l]{\hspace{0.05cm} ID4} & \makecell{14/14} & \makecell{18.35 $\pm$ 8.29} & \makecell{3.9 $\pm$ 2.3} & \makecell{0.10 $\pm$ 0.08}\\
        %\hline
        \makecell{chb04} & \makecell{3/4} & \makecell{\hspace{0.04cm} 23.09 $\pm$ 12.29} & \makecell{2.1 $\pm$ 1.5} & \makecell{0.16 $\pm$ 0.32} & \makecell[l]{\hspace{0.05cm} ID5} & \makecell{10/10} & \makecell{13.63 $\pm$ 8.30} & \makecell{4.5 $\pm$ 2.1} & \makecell{0.05 $\pm$ 0.09} \\
        %\hline
        
        \makecell{chb05} & \makecell{5/5} & \makecell{15.33 $\pm$ 9.32} & \makecell{2.3 $\pm$ 0.5} & \makecell{0} & \makecell[l]{\hspace{0.05cm} ID6} & \makecell{4/4} & \makecell{\hspace{0.04cm} 15.76 $\pm$ 10.56} & \makecell{5.5 $\pm$ 2.9} & \makecell{0.11 $\pm$ 0.13} \\
        %\hline
        
        \makecell{chb06} & \makecell{7/7} & \makecell{\hspace{0.04cm} 8.85 $\pm$ 5.22} & \makecell{2.3 $\pm$ 0.3} & \makecell{0} & \makecell[l]{\hspace{0.05cm} ID9} & \makecell{9/9} & \makecell{16.42 $\pm$ 9.53} & \makecell{3.1 $\pm$ 1.6} & \makecell{0.07 $\pm$ 0.11} \\
        
        \makecell{chb07} & \makecell{3/3} & \makecell{11.57 $\pm$ 3.71} & \makecell{2.7 $\pm$ 0.1} & \makecell{0} & \makecell[l]{\hspace{0.05cm} ID10} & \makecell{4/5} & \makecell{\hspace{0.04cm} 10.31 $\pm$ 10.46} & \makecell{4.7 $\pm$ 0.5} & \makecell{0.02 $\pm$ 0.05} \\
        %\hline
        
        \makecell{chb08} & \makecell{4/5} & \makecell{\hspace{0.04cm} 20.44 $\pm$ 14.72} & \makecell{2.5 $\pm$ 0.6} & \makecell{0.16 $\pm$ 0.26} & \makecell[l]{\hspace{0.05cm} ID12} & \makecell{10/10} & \makecell{12.86 $\pm$ 7.70} & \makecell{5.2 $\pm$ 1.5} & \makecell{0.01 $\pm$ 0.03} \\
        %\hline
        
        \makecell{chb09} & \makecell{4/4} & \makecell{11.31 $\pm$ 7.51} & \makecell{2.1 $\pm$ 0.6} & \makecell{0.30 $\pm$ 0.60} & \makecell[l]{\hspace{0.05cm} ID13} & \makecell{6/7} & \makecell{20.96 $\pm$ 8.53} & \makecell{6.2 $\pm$ 2.5} & \makecell{0.08 $\pm$ 0.05} \\
        %\hline
        
        \makecell{chb10} & \makecell{7/7} & \makecell{\hspace{0.04cm} 18.80 $\pm$ 10.19} & \makecell{2.4 $\pm$ 0.9} & \makecell{0.09 $\pm$ 0.24} & \makecell[l]{\hspace{0.05cm} ID14} & \makecell{5/7} & \makecell{28.07 $\pm$ 5.95} & \makecell{7.3 $\pm$ 3.0} & \makecell{0.21 $\pm$ 0.08} \\
        %\hline
        
        \makecell{chb11} & \makecell{3/3} & \makecell{16.82 $\pm$ 4.59} & \makecell{1.7 $\pm$ 0.6} & \makecell{0.10 $\pm$ 0.16} & \makecell[l]{\hspace{0.05cm} ID16} & \makecell{5/6} & \makecell{15.75 $\pm$ 6.44} & \makecell{4.6 $\pm$ 3.0} & \makecell{0.04 $\pm$ 0.06} \\
        
        % \cline{6-10}
        \makecell{chb14} & \makecell{8/8} & \makecell{10.46 $\pm$ 3.97} & \makecell{2.4 $\pm$ 0.3} & \makecell{0} & \makecell{} & \makecell{} & \makecell{} & \makecell{} & \makecell{}\\
        % \cline{6-10}
        
        \makecell{chb17} & \makecell{2/3} & \makecell{\hspace{0.04cm} 29.86 $\pm$ 14.90} & \makecell{3.2 $\pm$ 1.6} & \makecell{0} & \makecell{} & \makecell{} & \makecell{} & \makecell{} & \makecell{}\\
        
        \makecell{chb18} & \makecell{5/6} & \makecell{\hspace{0.04cm} 16.05 $\pm$ 15.51} & \makecell{2.5 $\pm$ 0.2} & \makecell{0} & \makecell{} & \makecell{} & \makecell{} & \makecell{} & \makecell{}\\
        
        \makecell{chb19} & \makecell{3/3} & \makecell{\hspace{0.04cm} 9.03 $\pm$ 5.11} & \makecell{2.5 $\pm$ 0.4} & \makecell{0} & \makecell{} & \makecell{} & \makecell{} & \makecell{} & \makecell{}\\
        
        \makecell{chb20} & \makecell{8/8} & \makecell{10.15 $\pm$ 6.33} & \makecell{1.9 $\pm$ 0.5} & \makecell{0.14 $\pm$ 0.40} & \makecell{} & \makecell{} & \makecell{} & \makecell{} & \makecell{}\\
        
        \makecell{chb21} & \makecell{4/4} & \makecell{19.93 $\pm$ 8.45} & \makecell{3.2 $\pm$ 0.4} & \makecell{0} & \makecell{} & \makecell{} & \makecell{} & \makecell{} & \makecell{}\\
        
        \makecell{chb22} & \makecell{3/3} & \makecell{ 23.26 $\pm$ 4.01} & \makecell{2.7 $\pm$ 1.2} & \makecell{0.24 $\pm$ 0.38} & \makecell{} & \makecell{} & \makecell{} & \makecell{} & \makecell{}\\
        
        \makecell{chb23} & \makecell{7/7} & \makecell{15.60 $\pm$ 7.70} & \makecell{2.1 $\pm$ 0.9} & \makecell{0} & \makecell{} & \makecell{} & \makecell{} & \makecell{} & \makecell{}\\
        \midrule
        % \cline{1-5}
        \makecell{Overall} & \makecell{94/99} & \makecell{14.84 $\pm$ 9.80} & \makecell{2.3 $\pm$ 0.7} & \makecell{0.08 $\pm$ 0.14} & \makecell{Overall} & \makecell{84/89} & \makecell{16.17 $\pm$ 9.26} & \makecell{4.7 $\pm$ 2.0} & \makecell{0.08 $\pm$ 0.09}\\
        \bottomrule

        \end{tabular}
        \begin{tablenotes}
            \scriptsize
            \item[1] RPIP = Rectified Predictive Ictal Probability \quad FDR = False Detection Rate
            \item[2] N$_{\text{DC}}$ = Number of Seizures Detected during the Crossing Period \quad N$_{\text{T}}$ = Number of Total Seizures
            
        \end{tablenotes}
        % \quad
        
        % \leftline{\text{\scriptsize{\hspace{0.5cm} RPIP: Rectified Predictive Ictal Probability \quad FDR: False Detection Rate}}}
        % \leftline{\text{\scriptsize{\hspace{0.5cm} N$_{\text{DC}}$: Number of Seizures Detected during the Crossing Period \quad N$_{\text{T}}$: Number of Total Seizures }}}
        % \leftline{\text{\scriptsize{}}}
        
\end{table*}

\section{Experiments}
\subsection{Experimental setting}
The experiments were implemented by Python with Pytorch framework, and M-3D-CNN model training and inference works are carried out on the single NVIDIA 2080Ti GPU machine. In this work, we trained patient-specific model training and implemented the leave-one-seizure-out cross-validation (LOSOCV) scheme, which means we select one seizure for validation, and the rest seizures are used to train the model. LOSOCV is meaningful from clinical perspective because the selected validated seizure can be regarded as a fresh seizure unseen by the model yet, if the model performs well in this scheme, we can believe that the trained model can also accurately and promptly alarm seizures in the future.

Experimentally, all interictal, ictal and crossing samples are used to train the model, and only errors of crossing samples are considered as the most crucial metric to quantify the quality of the model because the trained model can perfectly recognize accurate probabilities of complete interictal and ictal samples. During the phase of model training, we set 20 training epochs and used optimizer is Nesterov-accelerated Adam, known as Nadam, with 0.0001 learning rate, $\beta_{1}=0.9$, $\beta_{2}=0.999$ (\cite{nadam}). We implement the LOSOCV scheme and only save the best model performing the lowest RPIP errors of crossing samples on the validated seizure. In terms of rectified weighting strategy, weights $\lambda_{1}, \lambda_{2}, \lambda_{3}, \lambda_{4}$ are experimentally set to $0.2, 0.3, 0.3, 0.2$. The detection rate $r$ and decision threshold $Thr.$ are experimentally set to 10 and 0.5 for both datasets.

\begin{figure*}[!t]
\centerline{\includegraphics[width=\textwidth]{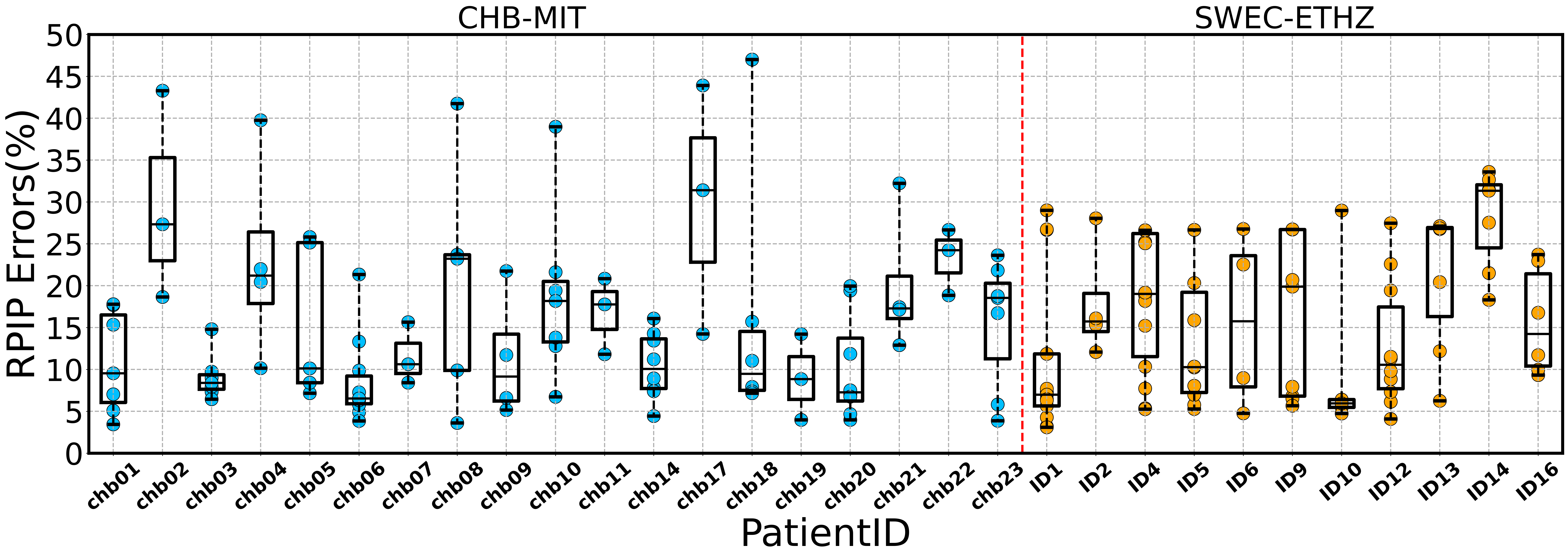}}
\caption{Boxplot for rectified predictive ictal probability of crossing samples in each seizure on two datasets. (RPIP = Rectified Predictive Ictal Probability)}
\label{rpip}
\end{figure*}

\begin{figure*}[!t]
\centerline{\includegraphics[width=\textwidth]{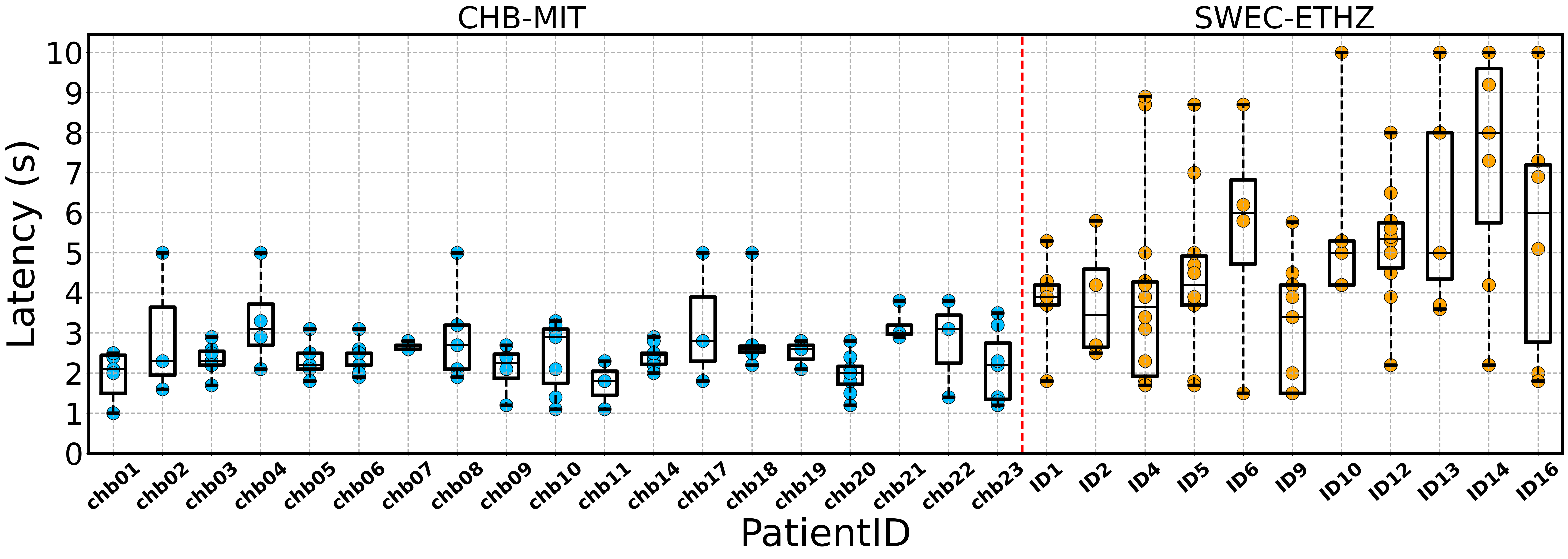}}
\caption{Boxplot for detection latency of each seizure on two datasets. The averaged latencies of two datasets are 2.3 s and 4.2 s, respectively.}
\label{late}
\end{figure*}

\subsection{Results}
At first, we need prove the effectiveness of rectified weighting strategy. Fig. \ref{rws} shows 3 representative examples of PIP and RPIP performance from CHB-MIT dataset, where red, blue, and orange dots stand for true, predictive, and rectified probabilities, respectively. The x-axis represents percentage of ictal period in crossing sample, and y-axis refers to probability. As mentioned in Section III-B and III-C, there are 1280 extracted crossing samples in crossing period for each seizure, and there are divided into 20 probability pairs annotation as true labels. Thus, every 5\% ictal period contains 64 samples. We can see that even though original PIPs perform well according to Fig. \ref{rws}(a), rectified weighting strategy still can enhance the results from 6\% to 4\%. In Fig. \ref{rws}(b), original PIPs show worse fitting result with 23\% errors, while rectified weighting strategy can significantly decrease the errors to 7\%, and the overall probabilities are increasing more linearly. Fig. \ref{rws}(c) is another type of representative example that RPIPs seem to achieve increased errors compared to original PIPs (from 7\% to 9\%), but we can see that RPIPs increase more linearly than PIPs, so that we still keep the results of RPIPs. According to these three representative examples, we can conclude that rectified weighting strategy is effective to rectify the PIPs by decreasing errors further and making PIPs increase more linearly as expected. This operation aims to help detection system can recognize samples more accurately and meanwhile decrease FDR.

Table 2 shows performances of proposed M-3D-CNN model on CHB-MIT and SWEC-ETHZ datasets. In this table, we only consider RPIPs after implementing rectified weighting strategy, then compute detection latency and FDR based on RPIPs. We achieved overall 94 of 99 and 84 of 89 seizures detected during the crossing period, 14.84\% $\pm$ 9.80\% and 16.17\% $\pm$ 9.26\% RPIP errors of crossing samples, 2.3 s $\pm$ 0.7 s and 4.7 s $\pm$ 2.0 s detection latency and 0.08/h $\pm$ 0.14/h and 0.08/h $\pm$ 0.09/h FDR on CHB-MIT scalp dataset and SWEC-ETHZ iEEG dataset, respectively. And 100\% seizures are detected after EEG onset for both datasets. These mean and standard deviation values are calculated based on results attained from various numbers of seizures in each patient according to LOSOCV scheme. Here we only display RPIPs of crossing samples, because interictal and inter samples can be accurately predicted by the model where RPIPs are <3\%, and detection latency significantly depends on RPIPs of crossing samples.

In terms of performance on CHB-MIT dataset, M-3D-CNN model performs well on 8 cases (chb01, chb03, chb05, chb06, chb07, chb14, chb19, chb23), all seizures are detected during the crossing period, low RPIP errors of crossing samples ($\leq$15\%), short detection latency ($\leq$2.7 s) and none false detection are attained on these patients. The rest subjects show slight drawbacks on one or two performance metrics. There are 5 patients (chb02, chb04, chb08, chb17, chb18) containing 1 seizure do not be detected by M-3D-CNN model during the crossing period. However, these miss-detected seizures still can be detected after crossing period, so that we set the detection latency of them to 5 s and 10 s which equals to the length of crossing period for CHB-MIT and SWEC-ETHZ datasets, respectively. We can see from table that 1 miss-detected seizure leads to high RPIP and FDR, except for chb18, M-3D-CNN model obtains larger than 20\% even close to 30\% mean and larger than 10\% standard deviation RPIP errors, and larger than 0.8/h FDR on the rest 4 patients (chb02, chb04, chb08, chb17). As for chb22 patient, even though there is no miss-detected seizure, we still achieved slightly higher RPIP errors and FDR. In terms of SWEC-ETHZ dataset, except for ID13 and ID14 subjects, proposed model performs well on the rest 8 patients, where $\leq$20\% RPIP errors of crossing samples and $\leq$5 s detection latency are achieved. There are 4 patients (ID10, ID13, ID14, ID16) containing miss-detected seizures during the crossing period, correspondingly worse performance metrics are achieved on these patients. Especially for ID14 case where M-3D-CNN model performs worst, there are two miss-detected seizures and the highest RPIP errors, detection latency, and FDR are attained.

\begin{table*}[!ht]
    \scriptsize
    \caption{Performance comparison between this work and prior-art studies.}
    \centering
    \renewcommand\arraystretch{2}
        \begin{tabular}[]{c|c|c|c|c|c|c|c|c|c}
        \toprule
         \makecell{Ref.} & \makecell{Dataset} & \makecell{EEG \\ type} & \makecell{Feature \\ extraction \\ method} & \makecell{Len. of \\ Sample} & \makecell{Model} & \makecell{FDR \\ (/h)} & \makecell{Detection \\ Latency} & \makecell{LOSO- \\ CV \\ scheme} & \makecell{Probabi- \\ listic \\ task} \\
        \midrule
        
        \makecell{\cite{ShoebG10}} & \makecell{CHB-MIT} & \makecell{EEG} & \makecell{FFT} & \makecell{6s} & \makecell{SVM} & \makecell{0.08} & \makecell{4.6s} & \makecell{\checkmark} & \makecell{--}\\
        % \hline
         \makecell{\cite{KHARBOUCH2011S29}} & \makecell{Clinical} & \makecell{iEEG} & \makecell{FFT} & \makecell{3s} & \makecell{SVM} & \makecell{0.03} & \makecell{5s} & \makecell{\checkmark} & \makecell{--}\\

        \makecell{\cite{7727334}} & \makecell{CHB-MIT} & \makecell{EEG} & \makecell{Raw} & \makecell{1s} & \makecell{RNN} & \makecell{0.08} & \makecell{7s} & \makecell{\checkmark} & \makecell{--}\\
        
        % \makecell{\cite{7911296}} &\makecell{CHB-MIT} & \makecell{EEG} &  \makecell{Wavelet} & \makecell{6s} & \makecell{SVM} & \makecell{96} & \makecell{0.1} & \makecell{1.89s (+6s)} & \makecell{\checkmark} & \makecell{--}\\
        
         % \makecell{\cite{7927402}} & \makecell{CHB-MIT} & \makecell{EEG} & \makecell{22} & \makecell{Statistical} & \makecell{6s} & \makecell{ADCD} & \makecell{96} & \makecell{0.12} & \makecell{4.21s (+3s)} & \makecell{\checkmark} & \makecell{--}\\
        
         \makecell{\cite{8470079}} &\makecell{CHB-MIT} & \makecell{EEG} & \makecell{STFT} & \makecell{3s} & \makecell{2D-CNN} & \makecell{--} & \makecell{--} & \makecell{--} & \makecell{--}\\
        
         \makecell{\cite{10.1145/3241056}} & \makecell{CHB-MIT} & \makecell{EEG} & \makecell{Raw} & \makecell{2s} & \makecell{2D-CNN} & \makecell{--} & \makecell{--} & \makecell{--} & \makecell{--} \\
        
         %\makecell{\cite{8995501}} & \makecell{CHB-MIT} & \makecell{EEG} & \makecell{Wavelet} & \makecell{--} & \makecell{2D-CNN} & \makecell{--} & \makecell{--} & \makecell{--} & \makecell{--} \\
        
         \makecell{\cite{8723166}} & \makecell{SWEC-ETHZ} & \makecell{iEEG} & \makecell{LBP \\ +  LGP} & \makecell{6-bit} & \makecell{SVM \\ MLP} & \makecell{0} & \makecell{15.9s} & \makecell{--} & \makecell{--} \\
        
         \makecell{\cite{9339990}} & \makecell{CHB-MIT} & \makecell{EEG} & \makecell{Wavelet \\ + EMD} & \makecell{4s} & \makecell{SVM} & \makecell{0.64} & \makecell{--} & \makecell{--} & \makecell{--} \\
        
         \makecell{\cite{WANG2021212}} & \makecell{CHB-MIT\\ SWEC-ETHZ} & \makecell{EEG\\ iEEG} & \makecell{Raw} & \makecell{2s} & \makecell{1D-CNN} & \makecell{0.2 \\ 0.07} & \makecell{8.1s \\ 13.2s} & \makecell{--} & \makecell{--} \\
         
         \makecell{\cite{9187540}} & \makecell{SWEC-ETHZ} & \makecell{iEEG} & \makecell{Statistical} & \makecell{2s} & \makecell{1D-CNN} & \makecell{--} & \makecell{8.8s} & \makecell{--} & \makecell{--} \\

         \makecell{\cite{SHEN2022103820}} & \makecell{CHB-MIT} & \makecell{EEG} & \makecell{Wavelet} & \makecell{30s} & \makecell{RUSBoosted} & \makecell{3.24} & \makecell{10.42s} & \makecell{--} & \makecell{--} \\
        
         \makecell{\cite{CIMR2023107277}} & \makecell{CHB-MIT} & \makecell{EEG} & \makecell{Raw} & \makecell{23.6s} & \makecell{2D-CNN}  & \makecell{3.11} & \makecell{--} & \makecell{--} & \makecell{--} \\

        \makecell{\cite{SHEN2023104566}} & \makecell{CHB-MIT} & \makecell{EEG} &  \makecell{Wavelet} & \makecell{5s} & \makecell{2D-CNN} & \makecell{2.1} & \makecell{10.46s} & \makecell{--} & \makecell{--} \\
        \midrule
        
        \makecell{\textbf{This work (2023)}} & \makecell{\textbf{CHB-MIT}\\\textbf{SWEC-ETHZ}} & \makecell{\textbf{EEG}\\ \textbf{iEEG}} & \makecell{\textbf{Multiscale} \\ \textbf{STFT}} & \makecell{\textbf{5s} \\ \textbf{10s}} & \makecell{\textbf{3D-CNN}} & \makecell{\textbf{0.08} \\ \textbf{0.08}} & \makecell{\textbf{2.3s} \\ \textbf{4.7s}} & \makecell{\checkmark} & \makecell{\checkmark}\\

        \bottomrule
        \end{tabular}
        \begin{tablenotes}
            \scriptsize
            \item[1] -- = N/A  \quad Len. = Length \quad FDR = False Detection Rate \quad LOSOCV = Leave-One-Seizure-Out Cross-Validation
            \item[2] iEEG = intracranial EEG \quad FFT = Fast Fourier Transform \quad EMD = Empirical Mode Decomposition
            \item[3] STFT = Short-Time Fourier Transform \quad LBP = Local Binary Pattern \quad LGP = Local Gradient Pattern
        \end{tablenotes}
        % \quad

        % % \leftline{\text{\scriptsize{\hspace{0.5cm} *-: N/A}}}
        % \leftline{\text{\scriptsize{\hspace{0.5cm} --: N/A  \hspace{0.65cm} Len.: Length \hspace{0.65cm} Sen.: Sensitivity \hspace{0.93cm} FDR: False Detection Rate \quad \quad \quad LOSOCV: Leave-One-Seizure-Out Cross Validation}}}
        % \leftline{\text{\scriptsize{\hspace{0.5cm} sEEG: scalp EEG \quad iEEG: intracranial EEG \quad FFT: Fast Fourier Transform \quad SVM: Support Vector Machine}}}
        % \leftline{\text{\scriptsize{\hspace{0.5cm} STFT: Short-Time Fourier Transform \hspace{0.9cm} ADCD: Adaptive Distance-based Change Point Detector}}}
        % \leftline{\text{\scriptsize{\hspace{0.5cm} LBP: Local Binary Pattern \quad LGP: Local Gradient Pattern \quad MLP: Multilayer Perception \quad EMD: Empirical Mode Decomposition}}} 
\end{table*}

Fig. \ref{rpip} and Fig. \ref{late} display boxplots specifying every seizure performance of RPIP errors and detection latency on two datasets, respectively. According to Fig. 7, it is obvious that most seizures achieved less than 30\% RPIP errors and the majority is less than 20\%, meanwhile there only 8 out of 99 seizures achieved larger than 30\%, and 4 of them achieved larger than 40\% on CHB-MIT dataset. Each of these seizures obtaining higher RPIP errors appears in different subjects, this phenomenon indicates that worse RPIP errors are not generated by the poor model or abnormal patient, this may be caused by a single abnormal seizure among these corresponding patients. And these possible abnormal seizures lead to a large standard deviation as shown in reverent cases from Table 2. In terms of SWEC-ETHZ dataset, all seizures obtained less than 30\% RPIP errors except for the aforementioned worst-performing case ID14, we suspect ID14 is an abnormal patient different from others. As for Fig. 8, we can see from CHB-MIT dataset that except for 5 miss-detected seizures during the crossing period set to 5 s latency, all rest seizures achieve less than 4 s latency, and the averaged latency is 2.3 s. From SWEC-ETHZ dataset, there are also 5 miss-detected seizures set to 10 s latency, among the rest seizures, around 10\% seizures are larger than 8 s, and the majority is less than 6s. The averaged detection latency among 89 seizures is 4.2 s.

Table 3 shows performance comparisons between prior-art publications and this work, we list several characteristics of the used dataset and proposed method to make comparisons. There are several previous highly cited works with prior-art performance selected to prove the advantages of our work. In should be noted that our defined event-based sensitivity is only applied to works in (\cite{ShoebG10,WANG2021212}), most sensitivity used in previous studies is sample-based sensitivity which cannot indicate clinical importance, so that we do not compare sensitivity metric.
%It should be noted that in terms of detection latency item, we add the length of detected sample to previously reported latencies because we doubt previous works did not compute the latency by measuring the distance between EEG onset and the end of detected sample if their latencies shorter than length of sample. 
According to prior publications, the state-of-the-art detection latencies on two datasets are 4.2 s and 8.1 s respectively. The latencies obtained by our proposed algorithm are 2.3 s and 4.7 s which are significantly faster than prior-art results. And we can see that several previous studies even if utilized the naive FFT feature extraction method and SVM classifier, they still achieved good performance. However, numerous recent emerging studies took advantage of advanced feature extraction methods and DL models, they cannot significantly enhance the seizure detection performance. Furthermore, less recent studies focused on testing meaningful metrics from clinical perspectives, such as FDR, detection latency, and LOSOCV scheme. In the last three columns of the table, we highlight three innovations of this work, which are whether or not implementing LOSOCV scheme and probabilistic prediction task.

\section{Discussion}
In this section, we discuss several issues, including further clarification of achieved performance, model comparison from hardware and performance perspectives.

\subsection{Performance clarification}
Firstly, we will clarify the sensitivity metric. In this work, we only consider the number of seizures detected during the crossing period as effectively detected seizures instead of seizures detected at any time as previous works did \cite{SHOEB2004483,ShoebG10}. Because we think the seizure can be detected during the crossing period means detection latency of this seizure is short enough to guarantee detection time can precede clinical onset. Experimentally, M-3D-CNN model can detect all seizures after crossing period (during the ictal period), so that sensitivity after EEG onset would be 100\% as the way previous researchers measured. But we think such 100\% sensitivity would not be clinically beneficial for epileptic patients.

Secondly, the lengths of extracted samples for two datasets are different. We empirically and experimentally set 5 s and 10 s for CHB-MIT and SWEC-ETHZ datasets, respectively. We initially learn from previous studies that achieved detection latency of two datasets were around 5 s and larger than 10 s. Then we experimentally tuned this parameter, we found that longer extracted samples would bring longer detection latency and lower FDR, while shorter extracted samples would cause shorter detection latency and higher FDR. Thus, this is a kind of trade-off parameter tuning work, eventually the length of extracted sample making the first ictal sample (or the last crossing sample) just contain the obvious EEG signal oscillations is expected, thereby we set 5 s and 10 s for two datasets.

Thirdly, achieved detection latency on SWEC-ETHZ is obviously longer than the latency achieved on CHB-MIT dataset. There are two possible reasons to answer this phenomenon. The first is that the criterion of annotating EEG onset time is implemented by different clinical experts. The second is that iEEG modality is more sensitive to the abnormal discharging inside brain, thereby can detect the slight EEG abnormality more earlier than scalp EEG \cite{BALL2009708}. However, such earlier slight abnormality appearing in EEG signal is difficult to be detected by the algorithm.

\begin{table}[!t]
    \scriptsize
    \caption{Performance comparisons of various Models on chb03 subject with 7 seizures from CHB-MIT dataset.}
    \centering
    \renewcommand\arraystretch{1.5}
        \begin{tabular}[]{c|c|c|c}
        \toprule
        \makecell{Model Name} & \makecell{PIP Errors of \\ Crossing Samples \\ (\%)} & \makecell{Number of \\ Parameters} & \makecell{Model \\ Size}  \\
        \midrule
        \makecell[r]{\textbf{M-3D-CNN}} & \makecell{\textbf{\hspace{0.04cm} 7.14 $\pm$ 3.38}} & \makecell{3.8M} & \makecell{14.46MB} \\
        % \makecell[r]{5-STFT-3D-CNN} & \makecell{\hspace{0.04cm} 9.72 $\pm$ 4.46} & \makecell{\hspace{0.1cm} 468,042} & \makecell{\hspace{0.05cm} 1.79MB} \\
        \makecell[r]{M-2D-CNN} & \makecell{12.41 $\pm$ 8.36} & \makecell{5.6M} & \makecell{21.42MB} \\
        \makecell[r]{5-2D-CNN} & \makecell{16.07 $\pm$ 5.34} & \makecell{1.2M} & \makecell{\hspace{0.05cm} 4.65MB} \\
        \makecell[r]{M-LSTM} & \makecell{\hspace{0.04cm} 8.53 $\pm$ 5.61} & \makecell{5.1M} & \makecell{19.56MB} \\
        \makecell[r]{5-LSTM} & \makecell{\hspace{0.04cm} 9.13 $\pm$ 5.20} & \makecell{5.1M} & \makecell{19.56MB} \\
        % \makecell[r]{Raw-2D-CNN} & \makecell{} & \makecell{} & \makecell{} \\
        \makecell[r]{M-ViT} & \makecell{\hspace{0.04cm} 8.87 $\pm$ 4.28} & \makecell{7.9M} & \makecell{30.32MB} \\
        \makecell[r]{5-ViT} & \makecell{\hspace{0.04cm} 9.65 $\pm$ 8.00} & \makecell{7.9M} & \makecell{30.32MB} \\
        % \makecell[r]{Raw-ViT} & \makecell{} & \makecell{} & \makecell{} \\

        % \makecell[r]{Raw-LSTM} & \makecell{} & \makecell{} & \makecell{} \\
        
        \bottomrule
        \end{tabular}
        \begin{tablenotes}
            \scriptsize
            \item[1] PIP = Predictive Ictal Probability \quad M = Multiscale \quad 5 = Scale 5
            \item[2] ViT = Vision Transformer \quad CNN = Convolutional Neural Networks
            \item[3] LSTM = Long Short-Term Memory

        \end{tablenotes}
        % \quad
        
        % \leftline{\text{\scriptsize{\hspace{0.4cm} PIP: Predictive Ictal Probability \quad M: Multiscale \quad 5: Scale 5}}}
        % \leftline{\text{\scriptsize{\hspace{0.4cm} ViT: Vision Transformer \quad CNN: Convolutional Neural Networks}}}
        % \leftline{\text{\scriptsize{\hspace{0.4cm} LSTM: Long Short-Term Memory}}}

\end{table}

\subsection{Model comparison}
In this manuscript, we proposed a novel M-3D-CNN model based on multiscale STFT features and 3D convolution operation in CNN backbone. In Table 4, we change three innovative parameters - multiscale or not, 3D or 2D, CNN or other DL backbone architectures, then generate several variant DL models to make comparisons with proposed M-3D-CNN model. The original PIP errors of crossing samples and the model size achieved on the chb03 patient from CHB-MIT dataset are used to compare the performance. According to the table, M-3D-CNN architecture performs best among all variant models. We can also conclude that multiscale is better than single-scale STFT features, and 3D-CNN outperforms 2D-CNN. Although 5-2D-CNN model shows advantages in model size, it obtains the worst PIP errors.
Furthermore, long short-term memory (LSTM) recurrent neural networks and vision transformer (ViT) also achieve satisfactory performance, but their model sizes are quite large compared to the M-3D-CNN model. ViT is an emerging and powerful DL model applied to various applications, but ViT model is too large to fit this real-time seizure detection application even if we already tried our best to shrink the model parameters, and eventually model size is still doubled as M-3D-CNN architecture. It should be noted that we flatten the time axis of all STFT features as various time steps for both LSTM and ViT models, which makes multiscale or single-scale features only change the length of time steps, and would not change the number of learnable parameters. 

\section{Conclusion}
The proposed framework uses a deep M-3D-CNN model intended to address several challenges and limitations in the field of seizure detection study. It consists of a novel probabilistic prediction concept to accurately recognize crossing samples. Then we also propose rectified weighting strategy and accumulative decision-making rule to significantly shorten the detection latency of seizure onset.

Furthermore, although proposed framework is intended for seizure detection application, the concept of probabilistic prediction, rectified weighting strategy and accumulative decision-making rule can be applied to other electrophysiological signal based real-time BCI applications. Also, it can benefit other detection systems to make decisions more promptly and accurately.

\section*{Acknowledgments}
The authors acknowledge start-up funds from Westlake University to the Center of Excellence in Biomedical Research on Advanced Integrated-on-chips Neurotechnologies (CenBRAIN Neurotech) for supporting this research project. This work was funded in part by the Zhejiang Key R\&D Program Project No. 2021C03002, and in part by the Zhejiang Leading Innovative and Entrepreneur Team Introduction Program No. 2020R01005.

% Use the style of numbering in square brackets.
% If nothing is used, default style will be taken.
%\begin{enumerate}[a)]
%\item 
%\item 
%\item 
%\end{enumerate}  

%\begin{itemize}
%\item 
%\item 
%\item 
%\end{itemize}  

%\begin{description}
%\item[]
%\item[] 
%\item[] 
%\end{description}  

% Uncomment and use as the case may be
%\begin{theorem} 
%\end{theorem}

% Uncomment and use as the case may be
%\begin{lemma} 
%\end{lemma}



\bibliographystyle{YX-elsarticle-harv}
\bibliography{YX-ref}

% % Here goes the biography details.

% % Here goes the biography details.

\end{document}